\documentclass[manuscript]{emulateapj}
\usepackage{times}

\slugcomment{}

\shorttitle{Dry merger rate at $<z> \sim 0.55$}
\shortauthors{De Propris et al.}

\begin{document}


\title{An upper limit to the dry merger rate at $<z> \sim 0.55$}


\author{Roberto De Propris\altaffilmark{1},
Simon P. Driver\altaffilmark{2},
Matthew Colless\altaffilmark{3},
Michael J. Drinkwater\altaffilmark{4},\\
Jon Loveday\altaffilmark{5},
Nicholas P. Ross\altaffilmark{6,7},
Joss Bland-Hawthorn\altaffilmark{3,8},
Donald G. York\altaffilmark{9},\\
Kevin Pimbblet\altaffilmark{4}}


\altaffiltext{1}{Cerro Tololo Inter-American Observatory, Casilla 603, La Serena, Chile}
\altaffiltext{2}{School of Physics and Astronomy, University of St. Andrews, North Haugh, 
St Andrews, Fife KY16 9SS, UK}
\altaffiltext{3}{Anglo-Australian Observatory, PO Box 296, Epping, NSW 1710, Australia}
\altaffiltext{4}{Department of Physics, University of Queensland, Brisbane, Queensland 4072, Australia}
\altaffiltext{5}{Astronomy Centre, University of Sussex, Falmer, Brighton, BN1 9QJ, UK}
\altaffiltext{6}{Physics Department, Durham University, South Road, Durham, DH1 3LE, UK}
\altaffiltext{7}{Department of Astronomy and Astrophysics, The Pennsylvania State University, 
525 Davey Laboratory, University Park, PA 16802, USA} 
\altaffiltext{8}{Institute of Astronomy, School of Physics, University of Sydney, NSW 2006, Australia}
\altaffiltext{9}{Department of Astronomy and Astrophysics, University of Chicago, 
5640 South Ellis Avenue, Chicago, IL 60637, USA}


\begin{abstract}

We measure the fraction of Luminous Red Galaxies (LRGs) in dynamically close 
pairs (with projected separation less than 20 $h^{-1}$ kpc and velocity difference
less than 500 km s$^{-1}$) to estimate the dry merger rate for galaxies with
$-23 < M(r)_{k+e,z=0.2} +5 \log h < -21.5$ and $0.45 < z < 0.65$ in the 2dF-SDSS LRG and 
QSO (2SLAQ) redshift survey. For galaxies with a luminosity ratio of $1:4$ or
greater we determine a $5\sigma$ upper limit to the merger fraction of 1.0\%
and a merger rate of $< 0.8 \times 10^{-5}$ Mpc$^{-3}$ Gyr$^{-1}$ (assuming
that all pairs merge on the shortest possible timescale set by dynamical
friction). This is significantly smaller than predicted by theoretical models
and suggests that major dry mergers do not contribute to the formation of the
red sequence at $z < 0.7$.

\end{abstract}





\section{Introduction}

In the standard $\Lambda$ Cold Dark Matter ($\Lambda$CDM) picture, galaxies are
assembled via the progressive (hierarchical) merger of increasingly more massive
subunits (see, e.g., \citealt{cole08,neistein08} for recent reviews). About 1/2
of the stellar mass in present-day massive ($L>L^*$) galaxies is expected to be
accreted via major mergers at $z < 1$ (e.g., \citealt{delucia06,delucia07}). Mergers
should therefore be a common occurrence in the life of galaxies and have a profound
influence on their properties (such as star formation, morphology and nuclear activity
among others).

Observational and theoretical evidence suggests that an increasing fraction of mergers
at lower redshifts should take place between spheroidal or gas-poor galaxies (`dry' mergers).
In their study of the morphology of merging pairs in a CDM universe, \cite{khochfar03,
khochfar05} found that elliptical galaxies brighter than $L^*$ were mainly formed via
major dry mergers and that gas-rich mergers were only important at lower luminosities.
Above a `threshold' mass of $\sim 6.3 \times 10^{10}$ M$_{\odot}$, galaxies are not
expected to grow their stellar mass by (induced) star formation, but their primary mode
of mass accretion at $z < 1$ is via gas-poor (stellar-dominated) mergers \citep{delucia06,
khochfar08}. Unlike gas-rich mergers, dry mergers are not expected to disturb the observed
tight scaling relations for early-type galaxies \citep{boylan05,boylan06} and to better
reproduce the internal structure of local ellipticals \citep{naab06,naab07}.

The first examples of dry mergers, in significant numbers, were observed as close pairs
of galaxies on the red sequence in deep images of high redshift clusters \citep{vandokkum99,
vandokkum01}, where approximately 50\% of galaxies were expected to have undergone a dry
merger to the present epoch. In the local universe, \cite{vandokkum05} estimated that $\sim 30\%$
of $z < 0.1$ bright ellipticals in the MUSYC and NDFWS surveys showed residual structural features 
indicative of a dry merger in the `recent' past.

\cite{bell06a} used close pairs in the Combo-17 survey having similar photometric redshifts {\it and} 
showing evidence of interactions to infer an integrated merger rate of $\sim 80\%$ since $z < 
0.8$ for red galaxies with $M_B < -20.5 + 5 \log h$, while \cite{lin08} used dynamically close 
pairs in the DEEP2 survey to derive an integrated dry merger rate of $24\%$ at $z < 1.2$ for 
galaxies with $-21 < M_B - 5 \log h < -19$. However, \cite{hsieh08} identify close pairs in the 
RCS survey and derive an integrated merger rate of only $6\%$ per Gyr since $z=0.8$ for galaxies 
with $-25 < M_r -5 \log h < -20$, and \cite{wen09} use the same approach as \cite{bell06a} on LRGs 
in the SDSS survey to determine a merger rate of $0.8\%$ per Gyr at $z < 0.12$ for galaxies with 
$M_r < -21.2 + 5 \log h$. \cite{bundy09} find evidence of few to zero pairs of red galaxies at $z < 
1.2$ in the GOODS data.

The small scale correlation function of LRGs in the SDSS at $z < 0.36$ analyzed by \cite{masjedi06,
masjedi08} is consistent with an upper limit of $< 1.7\%$ per Gyr to the dry merger rate for galaxies
with $M_i < -22.75 + 5 \log h$ at $0.16 < z < 0.30$. \cite{bell06b} used this method to derive
a merger rate of $4\%$ per Gyr for $M_B < -20.5 + 5 \log h$ galaxies at $0.4 < z < 0.8$. \cite{white07}
obtain an integrated merger rate of $\sim 30\%$ for LRGs at $0.5 < z < 0.9$ in the NDFWS survey.
Finally, \cite{wake08} apply the correlation function method to galaxies the 2SLAQ survey
to derive a merger rate of $2.4\%$ per Gyr at $0.19 < z < 0.55$.

The galaxy merger rate has been usually measured via the fraction of galaxies in
close pairs, often with the added requirement of closeness in velocity space to
cull interlopers (e.g., \citealt{patton00,patton02} -- hereafter, P00,P02), and 
the fraction of galaxies showing significant asymmetries in their light distribution
(e.g., \citealt{conselice03,conselice03b}). The rationale behind the former approach
is that if a merger is to occur, a companion must be present, and therefore the close
pair fraction is related to the {\it future} merger rate for the galaxies being
considered. The latter method relies on the observation that if a galaxy has undergone
a recent merger, it is likely to be morphologically disturbed, and therefore an
asymmetric light distribution would be a signpost of a recent merger. A critical
discussion of these approaches may be found in \cite{depropris07} and \cite{genel08}.

In this paper we derive the dynamically close pair fraction for galaxies in the 2SLAQ survey and 
determine an upper limit to their merger rate. The benefit of using close pairs is that 
we are able to select major merger candidates between galaxies in a specified mass range 
(by appropriately choosing the luminosity of and magnitude difference between participating 
galaxies), derive a merger rate within a specified timescale (dependent on the chosen 
projected and velocity separations for the pair members and theoretical simulations) and 
identify ongoing merger candidates for later study. Galaxy asymmetries tend to be more 
difficult to interpret in this fashion and usually require better quality imaging than we have 
available in our survey, especially for dry mergers \citep{bell06a,wen09}, to which the 
2SLAQ survey is most sensitive. Unlike asymmetries, galaxy pairs are sensitive to the merger 
fraction of `progenitor' halos and this quantity may be more directly compared to theoretical 
models \citep{genel08}. 

The structure of this paper is as follows: in the next section we present the data, describe
the methodology and derive the pair fraction and an upper limit to the dry merger rate at
the intermediate redshifts sampled by the 2SLAQ survey. We then discuss and compare our 
results in the context of galaxy formation models and recent work on the dry merger rate. 
We adopt the latest cosmological parameters with $\Omega_M=0.27$, $\Omega_{\Lambda}=0.73$ and 
$H_0=100$ km/s/Mpc. Unless otherwise stated, all absolute magnitudes quoted in the following
are intended as including a term of $+5\,\log h$ and all distance measures need to be referred
to $h$ (to the appropriate power).
 
\section{The 2SLAQ survey data}

The data used in the 2SLAQ survey consist of LRGs with $i < 19.8$ mag. from the original sample 
of \cite{eisenstein01}, selected by their $g-r$ and $r-i$ colors to lie at $0.45 < z < 0.65$ 
(see \citealt{fukugita96} for a description of the SDSS filter system). Photometry and astrometry
for the target LRGs are derived from the SDSS Data Release 1 \citep{york00,abazajian03} with 
improvements from the latest release available at the time of the spectroscopic observations
(DR4 -- \citealt{adelman06}). The objects lie in two long strips on the celestial equator, divided 
into several disconnected patches, each of which is between 10 and 30 deg$^2$ in area, for a total 
survey coverage of 182 deg$^2$. 

Spectroscopy for candidate LRGs was carried out at the Anglo-Australian telescope 
using the 2dF facility \citep{lewis02}. For objects in the main sample (Sample 8), it was 
found that most galaxies are within the specified redshift limits and were observed with 
high spectroscopic completeness (typically $87\%$). A complete description of the data can 
be found in the general survey paper by Cannon et al. (2006; hereafter C06). Because we have
highly complete spectroscopy, we can confirm that at least 90\% of our galaxies have K-type
or LRG spectra, with no sign of star formation, while less than 1\% of the sample shows
emission lines \citep{roseboom06}. We can therefore use this sample to measure the dry
merger rate.

Following \cite{wake06}, we computed the absolute magnitude for objects with reliable redshifts,
$k+e$ corrected to the SDSS $r$ band at $z=0.2$. This facilitates comparison with previous work 
on galaxy evolution and the merger rate from the SDSS (e.g., \citealt{masjedi06,masjedi08}). Note 
that no correction for internal extinction was applied to these galaxies.

\section{Methodology}

We calculate close pair statistics following the formalism developed by Patton et al. (2000,
2002; hereafter P00, P02) for the SSRS2 and CNOC2 surveys. Here, we give an `algorithmic' 
description of the procedures used, and show how we apply weights to correct for sources of 
incompleteness and the flux limited nature of the 2SLAQ survey. A fuller description of the 
method can be found in P00, P02.

\subsection{Luminosity of the pair sample}

Let us consider a sample of $N_1$ primary galaxies brighter than a limiting absolute 
magnitude $M_1$ in some volume of space, and, in the same volume, a sample of $N_2$ 
secondary galaxies brighter than a limiting absolute magnitude $M_2$. The two samples 
may coincide (i.e., $M_1=M_2$), as is often the case for redshift surveys: we then
study `major' mergers between galaxies of approximately similar luminosity. We are 
interested in knowing the fraction of galaxies in the secondary sample that are dynamically
close to galaxies in the primary sample. We define two galaxies to be dynamically close if 
they have a projected separation $r_p < 20\, h^{-1}$ kpc and a velocity difference of $< 500$ km 
s$^{-1}$, as used by P00, P02 and in subsequent work. 

Of course, pair statistics are usually computed from redshift surveys, which are flux-limited
rather than volume limited. We therefore need to account for the dependence of pair counts on
the clustering properties and the mean density of galaxies in the sample, and to correct for
sources of spatial and spectroscopic incompleteness. Because clustering is luminosity dependent 
(e.g., \citealt{norberg02}) we follow P00 and restrict the analysis to a fixed range in luminosity, 
within which clustering properties are not expected to vary significantly, by imposing additional 
bright ($M_{bright}$) and faint ($M_{faint}$) limits on the sample. This means that we derive a pair 
fraction (and merger rate) for galaxies within a specified range and ratio in luminosity.

In our case, we select galaxies with $0.45 < z < 0.65$ (where the survey is most complete) 
and with $M_{bright}=-23$ and $M_{faint}=-21.5$ mag. Because the range of luminosities we
survey is small, we let $M_{faint}$ coincide with $M_2$. We also let the primary and secondary 
samples coincide, to select major mergers between galaxies of approximately equal luminosities 
(and make full use of the spectroscopic sample). Figure 1 shows the distribution of 2SLAQ 
galaxies in absolute magnitude vs. redshift, together with selection lines in magnitude and 
redshift (see below for details). 

\begin{figure}
\epsscale{0.8}
\plotone{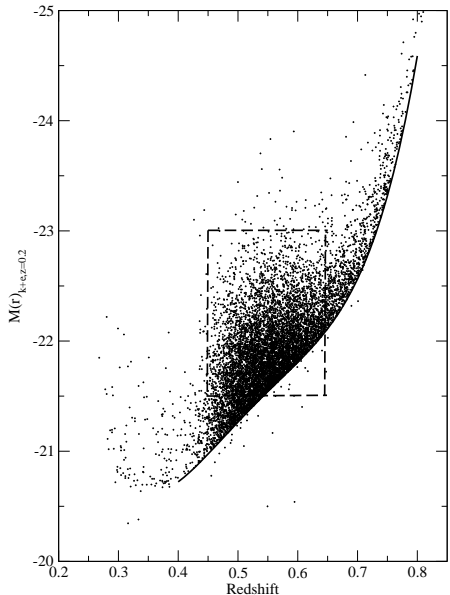}
\caption{The distribution of absolute magnitudes for 2SLAQ galaxies
vs. redshift. The thick solid line is the redshift-dependent absolute
magnitude limit for selection $M_{lim}(z)$ (see equation (1) and related
explanation in text). The thick dashed lines mark the ideal volume
limited box to which the observed sample is corrected using the
weights described in equations (2)-(9).}
\end{figure}

The faint limit allows us to include most 2SLAQ galaxies at $z > 0.45$ and reaches slightly 
below the luminosity of normal $L^*$ galaxies at the mean survey redshift (or, similarly, 
below the turnover in the 2SLAQ LRG luminosity function of \citealt{wake06}). We choose the 
bright limit of $M_r=-23$ to avoid the rarer and most luminous galaxies in 2SLAQ, which are 
likely to be more biased, and to mitigate the effects of luminosity-dependent clustering (as 
described above). With these choices, we study major mergers, with luminosity ratio $\geq 1:4$, where 
the secondary is at least 4 times less luminous than the primary galaxy\footnote{Note 
that of course no sample can be made complete for a given luminosity ratio without discarding a large 
fraction of the data (e.g., \citealt{patton08}), so theoretical comparisons need to take into account
the observational limitations of this technique.}, for $M_r < -21.5$. 

We can translate our luminosity range into a stellar mass range using the conversion between
$r$ band absolute magnitude and stellar mass by \cite{baldry06}, for typical LRG colors ($u-r >3$)
and with assumptions as to stellar populations as in \cite{baldry06}. With these choices
our $-23 < M_r < -21.5$ luminosity range corresponds to a stellar mass range of $16$--$5 \times 
10^{10}$ M$_{\odot}$.

\subsection{Density weighting $S(z)$}

The pair fraction that we actually measure needs to be corrected to the value that would 
be observed in an ideal volume-limited survey. In a flux-limited sample, primary galaxies 
at lower redshifts will have a greater likelihood of having a secondary companion within 
the survey than galaxies at higher redshift. We correct for this bias by assigning a greater 
weight to the rarer companions found at the high redshift end of the survey. This is carried 
out by computing for each galaxy a weight that renormalizes the sample to the density corresponding 
to a volume limited sample within $M_{bright} < M < M_2$. The weight is calculated by 
integrating the luminosity function over the appropriate ranges in absolute magnitude.

At each redshift, we then search for pairs of galaxies within $M_{bright}$ and a 
redshift-dependent absolute magnitude limit $M_{lim} (z)$, which is defined as:

\begin{equation}
M_{lim}(z) = {\rm max}[M_{faint},\, 19.8-5 \log d_L(z)-25-k(z)-e(z)]
\end{equation}

where $i=19.8$ mag. is the apparent magnitude limit of the survey, $d_L(z)$ is the luminosity
distance and $k(z)$ and $e(z)$ are the $k$ and evolutionary corrections. Recall that here
$M_{faint}$ is set to coincide with $M_2$. The $k+e$ corrections are taken to be the maximal 
corrections for a galaxy formed at high redshift and undergoing pure passive evolution, as 
other choices would allow galaxies to fall in and out of the sample according to their star 
formation histories (P00, P02). Note that here ${\rm max}$ means `the brightest of' rather 
than the (arithmetically) larger quantity. Fig.~1 shows this selection limit as applied to 
2SLAQ data.

Each secondary galaxy is weighted by the inverse of a selection function $S(z)$ 
defined as the ratio of densities in volume-limited vs. flux-limited samples:

\begin{equation}
S_N(z)={ {\int^{M_{lim}(z)}_{M_{bright}} \Phi(M) dM} \over
           {\int^{M_2}_{M_{bright}} \Phi(M) dM}}
\end{equation}

\begin{equation}
S_L(z)={ {\int^{M_{lim}(z)}_{M_{bright}} \Phi(M) L(M) dM } \over
           {\int^{M_2}_{M_{bright}} \Phi(M) L(M) dM}}
\end{equation}

where $L(M)=10^{0.4(M-M_{\odot})} L_{\odot}$ and $\Phi(M)$ is the LRG luminosity function
from \cite{wake06}. The integrals run from $M_{bright}$ to either the faint absolute limit 
$M_2$ (in the denominator) or to the redshift dependent absolute magnitude limit $M_{lim} (z)$
(in the numerator) set by the apparent magnitude limit of the survey.

Galaxies in the primary sample at low redshift will also have the largest number of
observed companions, while primaries at higher redshifts will have fewer observed
companions. This effect is corrected in a similar fashion as for galaxies in the
secondary sample, by applying the inverse of the secondaries' weight to primaries, 
i.e $S_N(z)$ and $S_L(z)$.

\subsection{Spatial incompleteness $w_b,w_v$}

We need to account for pairs missed because one of the galaxies falls outside of the
survey footprint. A potential companion may lie beyond the survey limits on the sky or 
be hidden in the `shadow' of a bright star, where galaxies cannot be detected or the 
spectra are contaminated\footnote{This is calculated using a formula developed by I. 
Strateva (unpublished) for the SDSS survey}. For each primary galaxy we compute
the fraction $1-f_b$ of the $\pi r_p^2$ area that may lie outside of the effective
survey area. The weight to be applied to the secondary galaxies is then $w_{b_2}=
1/f_b$. Primary galaxies may similarly be lost in the survey boundaries and the
appropriate weight to be applied to the primary sample is $w_{b_1}=f_b=w^{-1}_{b_2}$.

For objects with small separations, the SDSS photometric pipeline tends to merge
pairs into a single galaxy. For the LRG sample studied by \cite{masjedi06} the 
pipeline becomes unreliable for $r_p < 3''$. The $20$ $h^{-1}$ kpc search radius
we use corresponds to angular separations of $4.91''$ to $4.07''$ at $0.45 < z <
0.65$. We inspected all our 7889 images to verify whether the SDSS pipeline correctly
identifies photometric companions, using the SDSS 'Image List/Navigate' tools. The
results are that the pipeline certainly finds all targets with projected separations
between $3''$ and $5''$. For smaller separations the pipeline is more erratic, especially
for fainter systems. We therefore exclude an area of $3''$ around each target from our
search for companions and correct for the missed region via the $w_b$ weight, where
we assume that the distribution of pairs is uniform with projected separation over these
small scales. The size of the added factor lies between 37\% and 54\% depending on 
redshift.

The SDSS pipeline also tends to `double count' light for close pairs, making these
galaxies brighter than they would otherwise be. Because fainter galaxies have more 
mergers and minor mergers are more common \citep{patton08}, this has the effect of 
artificially raising the pair fraction. \cite{masjedi06} estimate that this should
decrease the actual merger rate for galaxies with $r_p < 5''$ by a factor of about
5. Since this is an uncertain correction and we are interested in an upper limit to
the merger rate, we do not consider this effect here (the effect cannot be well modelled
for our sample, as the SDSS pipeline cannot be run locally).

We also correct for possible companions missed by the redshift `cuts' we apply to the survey. 
If the primary lies within 500 km s$^{-1}$ of the redshift boundaries, we ignore all companions 
between the primary and the borders and apply a weight $w_{v_2}=2$ to all other companions in 
the opposite `direction'. We also need to apply a similar weight to account for potential primaries 
lost in the redshift boundary. This weight is the reciprocal of the weight applied to the 
secondaries, i.e., $w_{v_1}=1/2$

\subsection{Spectroscopic incompleteness $w_{\theta}$}

As all redshift surveys, 2SLAQ is not complete to its flux limit. Our success rate is $87\%$
for all galaxies within Sample 8 of C06 but the spectroscopic incompleteness may be dependent
on the separation between galaxies. Fibers in the 2dF positioner cannot be placed closer 
than about $25''$ from each other in every single configuration. However, this effect is 
compensated by the overlap between individual 2SLAQ tiles and by the fact that fiber 
configurations were generally retouched halfway through each (typically 4 hours) exposure 
to place fibers assigned to galaxies for which a reliable redshift had already been obtained 
on to a nearby target (see C06 for a description of the observations).

In order to correct for this source of bias we need to estimate the relative incompleteness
for close pairs over the range of separations of interest (corresponding to 20 $h^{-1}$ kpc)
and compare it with the incompleteness at large separations, where fiber interactions are not 
important. We follow P02 and estimate this weight by computing the ratio between the number of 
pairs between galaxies with redshift information ($N_{zz}$) {\it within the redshift range of
interest} and the number of pairs in the input photometric sample ($N_{pp}$), which is by definition 
complete, as a function of angular separation $\theta$ The weight to be applied is the ratio between the 
spectroscopic and photometric pair completeness at each separation normalized to the value at large 
separations.

In 2SLAQ we have higher overall completeness than CNOC2 (by about a factor of 2) but sample a smaller 
range of projected separations (because of our higher mean redshift). We plot $N_{zz} (\theta) /
N_{pp} (\theta)$ vs. $\theta$ in Fig.~2 for Sample 8 targets; the error bars are assumed to be
Poissonian. Although the data are noisy (because there are relatively few potential pairs to
start with), the value of $N_{zz}/N_{pp}$ at $\theta < 6''$ is consistent with the value of 
this ratio at $\theta > 100''$, arguing that we are not systematically more incomplete at small 
separations than at large ones where we are not affected by fiber collisions. 

P02 model the incompleteness in the CNOC2 survey by fitting a polynomial to the ratio 
$N_{zz}(\theta)/N_{pp}(\theta)$ as a function of projected separation $\theta$. Given the 
small number statistics and noisier nature of our data, it is not fully justified
to model the completeness as a function of $\theta$ with a polynomial as done in P02. We 
therefore adopt an uniform weight of 1 for 2SLAQ galaxies over the separation of interest,
as we do not appear to be systematically more incomplete than at larger separations. Although 
pair completeness drops at $30''$--$60''$ separations, because of fiber collisions, these large 
separations are not relevant to our study (as pairs with $r_p > 50$ $h^{-1}$ kpc or $\Delta v 
< 1000$ km $s^{-1}$ are largely spurious -- P00, De Propris et al. 2007).

\begin{figure}
\plotone{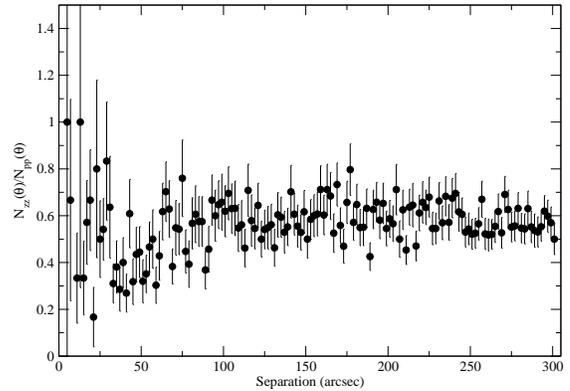}
\caption{The ratio of galaxies in pairs between galaxies with redshifts and
galaxies in pairs in the photometric sample, as a function of pair separation
$\theta$.}
\end{figure}

Given the small number statistics, and the fact we are ultimately deriving an upper limit,
we eventually decided to follow the approach by \cite{depropris05} to estimate the contribution 
from close pairs missed because of fiber collisions. We search around each of our main targets (Sample 8 galaxies with $-23.0 < M(r) 
< -21.5$ and  $0.45 < z < 0.65$) for a companion (lying within $r_p$) in the sample of galaxies 
(from the input photometric sample) for which we did not obtain a valid redshift. If such a companion 
exists, we assign to it the same redshift as the primary galaxy and require that the companion lies 
within the selection lines in Fig.~1. This identifies all pairs (a total of 3) which are potentially 
missed because of  redshift incompleteness. Pairs where both members are missed by the spectroscopic 
survey, will share in the general 2SLAQ incompleteness, without a bias for incompleteness at small 
angular separations. We can assume that all these 3 `extra' pairs are real and treat them as a source 
of systematic error on our determination of the pair fraction and the merger rate. Figure 3 shows 
postage stamp images of the two dynamical pairs we find and of the three possible pairs.

\begin{figure}
\plotone{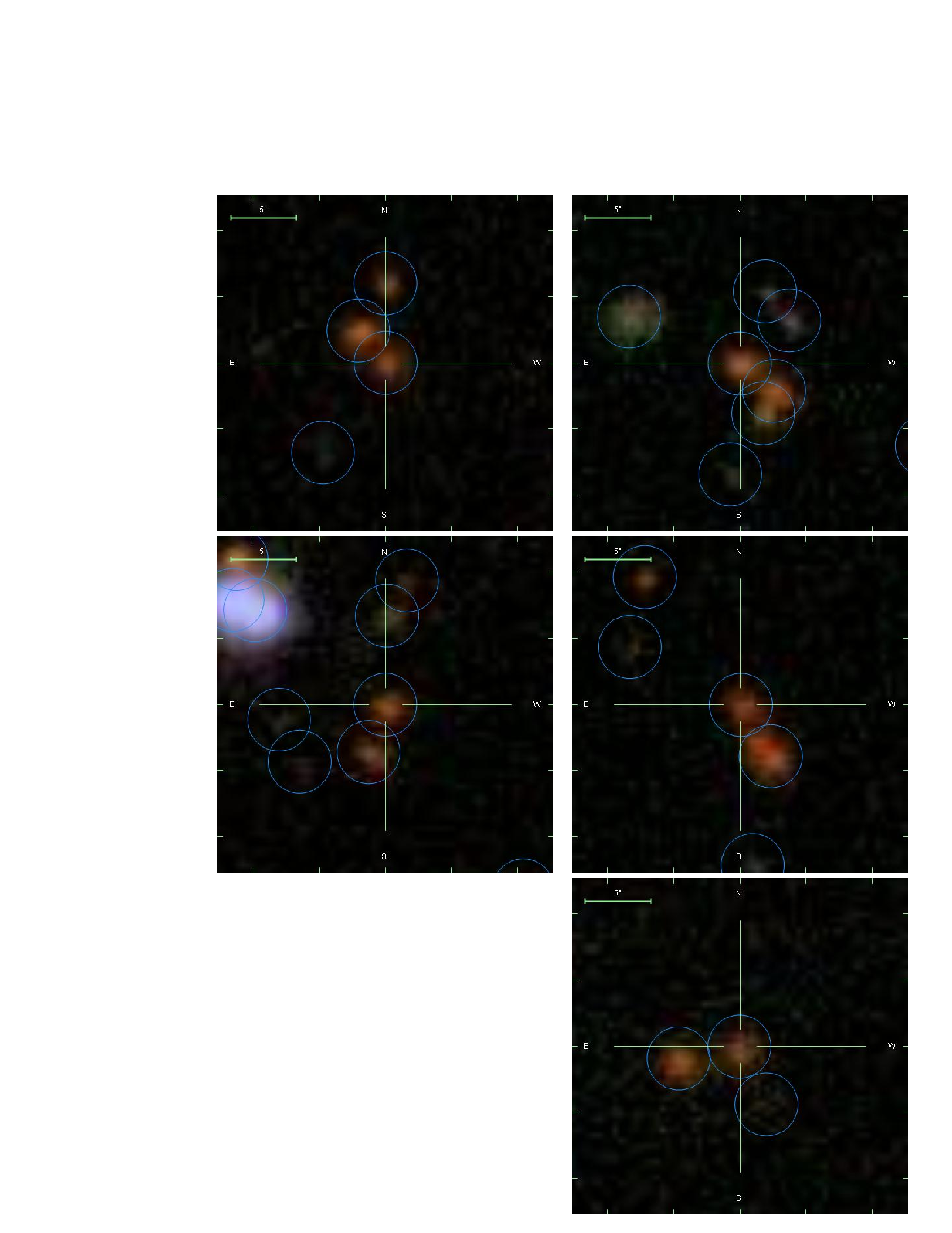}
\caption{Images (from the SDSS) of pairs found in our survey. The two left-hand
ones are the real ones, while the three images on the right hand show the `possible'
pairs}
\end{figure}

P02 also use a weight $w_s$ which accounts for the local magnitude incompleteness around each 
galaxy, a geometric effect due to slit placement and limiting filters in the CNOC2 survey, a 
color term and a term that depends on the evolution of the galaxy luminosity function
over the redshifts covered by the survey. This weight is specific to the methods employed by the
CNOC2 survey \citep{yee96}. In our case, we have a very homogeneous sample, there is no geometric
effect (other than the one corrected by $w_{\theta}$), all galaxies have similar colors and
the color selection is very efficient in selecting galaxies of the appropriate magnitude and redshift. 
Even in the CNOC2 survey most of these weights are equal to 1 \citep{yee96} for most galaxies. 
For these reasons, we do not adopt this weight in our study.

\subsection{Calculation of pair fraction}

Following P00 the number of close companions per galaxy in a flux-limited sample can be expressed as:

\begin{equation}
N_c={{\sum_i^{N_1} w^i_{N_1} N_{c_i}} \over {\sum_i^{N_1} w^i_{N_1}}}
\end{equation}

and the total companion luminosity:

\begin{equation}
L_c={{\sum_i^{L_1} w^i_{L_1} L_{c_i}} \over {\sum_i^{L_1} w^i_{L_1}}}\ L_{\odot}
\end{equation}

The sums run over the $i=1,...,N_1$ primary galaxies: $N_{c_i}$ and $L_{c_i}$ are 
the number and summed luminosity of galaxies from the secondary sample that are 
dynamically close (as defined above) to the  $i^{th}$ primary galaxy and are expressed as:

\begin{equation}
N_{c_i}=\sum_j w^j_{N_2}={\sum_j {w_{b_2}^j w^j_{v_2}} \over {S_N(z_j)}}
\end{equation}

\begin{equation}
L_{c_i}=\sum_j w^j_{N_2} L_j =\sum_j {{w_{b_2}^j w^j_{v_2}} \over {S_L(z_j)}} L_j
\end{equation}

where the sums run over those $j=1,...,N_2$ galaxies in the secondary sample that fulfill the criteria
for being dynamically close to the $i^{th}$ galaxy in the primary sample. 

The weights to be applied to the secondary sample correct for spatial incompleteness ($w^j_b$,$w^j_v$) 
and the flux-limit of the survey ($S(z_j)$) for each $j^{th}$ companion. We do not apply the spectroscopic
incompleteness weight $w_{\theta}$ because we use the alternate method described above to calculate the
contribution from missed close pairs. The weights are defined in the previous subsections. Similar weights 
also need to be applied to the primary sample, to correct for spatial incompleteness and the mean density:

\begin{equation}
w^i_{N_1}=w^i_{b_1} w^i_{v_1} S_N(z_i)
\end{equation}

\begin{equation}
w^i_{L_1}=w^i_{b_1} w^i_{v_1} S_L(z_i)
\end{equation}

where $w^i_{b_1}=f_b^i$ and $w^i_{v_1}=1/2$ (i.e., the reciprocals of the weights applied to the
secondary sample). No spectroscopic completeness weight is needed for the primary sample.

\subsection{Merger timescales}

In order to convert a pair fraction into a merger rate the timescale for the merger event needs
to be estimated. Merger timescales are difficult to determine and depend on the poorly known details 
of the merger process and how the potential of the individual galaxies reacts to the merger episode. 
The more commonly used timescales in previous work and semi-analytic models are based on dynamical 
friction arguments. The dynamical friction timescale can be calculated as \citep{patton00}:

\begin{equation}
T_{fric}={{2.64 \times 10^5\,r^2\,v_c} \over {M \ln \Lambda}}
\end{equation}

where r is initial physical pair separation in kpc, $v_c$ is the circular velocity in km s$^{-1}$,
$M$ is the mass and $\Lambda$ is the Coulomb logarithm. Assuming $r=20\,h^-1$ kpc, $v_c=260$ km s$^{-1}$
and $\ln \Lambda=2$ for equal mass mergers, we get a typical dynamical friction timescale between 0.1
and 0.3 Gyrs over the range of masses we sample.

The calibration by \cite{kitzbichler08} for the merger timescale of pairs in N-body simulations with 
$M > 5 \times 10^9$ M$_{\odot}$ and a velocity separation of 300 km s$^{-1}$ yields 0.9 Gyr, while 
the estimated merger timescale from the N-body simulations of \cite{boylan08} is $\sim 0.8$ Gyr for
the mass ratios we consider. A comparison between merger timescale by dynamical friction and in
N-body simulations by \cite{boylan08} shows that the dynamical friction formula underestimates the
merger timescale by a factor between 1.7 and 3.3 for mergers of mass ratio 1:3 to 1:10 respectively,
and leads to a 40\% overestimate of the mass accretion rate (mainly via minor mergers).

\section{Discussion}

\subsection{Results}

We can now apply the above procedure to the 2SLAQ sample. Out of 7889 galaxies we find a total
of one dynamically close pair (we find a second pair, but it lies within the $3''$ exclusion
circle). This yields a pair fraction $N_c=0.047\%$ and a luminosity accretion rate $L_c=2.5 
\times 10^7$ $L_{\odot}$. For galaxies within $z < 0.55$ this is $N_c=0.041\%$. Normally, 
errors on these quantities can be estimated by bootstrap resampling or jack-knifing, but 
this is not feasible with a sample of only two objects. We then proceed to estimate an 
upper limit to the pair fraction and the merger rate.

In order to derive an upper limit to the merger rate, we take the observed detections and use the 
binomial distribution. For large samples ($N > 100$) the error distribution is given by \citep{burgasser03}:

\begin{equation}
(\epsilon^U_b - \epsilon_b)/\epsilon_b=(\epsilon_b-\epsilon^L_b)/\epsilon_b=
\sqrt{1/n+1/N}
\end{equation}

where $\epsilon_b$ is the pair fraction, $n$ is the number of pairs, $N$ is the number
of objects in the sample, $\epsilon^U_b$ is the upper $1\sigma$ probability limit to the 
pair fraction and $\epsilon^L_b$ is the lower $1\sigma$ probability limit to the pair fraction.
The region between $\epsilon^U_b$ and $\epsilon^L_b$ corresponds to the 68\% confidence interval
for $\epsilon_b$ for a Gaussian distribution. In order to obtain aconservative upper limit we 
decide to use both pairs that we actually find (even though one of these is not within the
valid range of separations). 

For $n=2$ and $N=7889$ this yields $N_c=0.094 \pm 0.066\,\%$. The systematic error due to the 
three possible extra pairs is $0.14 \pm 0.08\%$. We can therefore quote a 5$\sigma$ upper limit 
to the pair fraction of $0.44\%$ (random) plus $0.54\%$ (systematic), for a total of $1.0\%$.

As a check, we estimate the pair fraction in each of the separate 2SLAQ survey patches: while 
most regions have no pairs, in one area we find a pair fraction of $0.30\%$, which is consistent 
with the upper limit we derived above. Finally, we can also consider pairs with wider separation 
in both projected distance and velocity: $r_p < 50\, h^{-1}$ kpc and $\Delta v < 1000$ km s$^{-1}$. 
This considerably increases the number of contaminants (unphysical pairs), as shown by P00 and 
\cite{depropris07}. However, this pair fraction may provide a further interesting constraint. 
For the 2SLAQ sample we find $N_c=0.13\%$. 

Our upper limit to the pair fraction can be translated into a $5\sigma$ upper limit to the dry 
merger rate using the expression:

\begin{equation}
R_{mg}=N_c\,n(z)\,0.5\,p_{merg}\,T^{-1}_{mg}
\end{equation}

where $n(z)$ is the space density of galaxies in our sample, 0.5 is a factor introduced 
to avoid double counting galaxies (1 pair contains 2 galaxies), $p_{merg}$ is the probability 
that pairs will merge (assumed to be 1 here) and $T_{mg}$ is the merger timescale, for which 
we take the shortest possible timescale set by dynamical friction. We find that a robust $5\sigma$ 
upper limit to the merger rate is: $ < 0.8 \times 10^{-5}$ Mpc$^{-3}$ $h^{-3}$ Gyr$^{-1}$ (including 
the systematic contribution due to the three extra photometric pairs) for galaxies with $-23 < M_r 
< -21.5$ at $0.45 < z < 0.65$ (a period of 1.4 Gyrs in the history of the Universe). A more realistic merger timescale and merger fraction may decrease this estimate by more than one order of magnitude.

\subsection{Comparison with Previous Measurements}

The low dry merger rate we measure here is in good agreement with a number of other
estimates: locally, \cite{masjedi06,masjedi08} obtain an upper limit of $< 1.7\%$ per
Gyr for the dry merger rate of SDSS LRGs with $M_i < -22.75$ and $z < 0.36$, a value
which is consistent with ours, albeit for more luminous (massive) galaxies. This is
also similar to the dry merger rate of $0.8\%$ measured by \cite{wen09} for SDSS
LRGs with $M_r < -21.5$ and $z < 0.12$. The integrated dry merger rate for galaxies
in the Red Sequence Survey is $\sim 6\%$ since $z=0.8$ over $-25 < M_r < -20$, while
\cite{bundy09} find very low to zero likely dry mergers in GOODS data at $z < 1.2$.

Nevertheless there are some discrepant estimates in the literature. Apparently, the
most worrying is the value of $2.4\%$ per Gyr for $0.19 < z < 0.55$ LRGs from 2SLAQ
derived by \cite{wake08} using the small scale correlation function. However, \cite{
wake08} use LRGs over the entire range of absolute luminosities in the 2SLAQ survey,
and their sample therefore includes both minor mergers (luminosity ratio greater 
than 1:4) and less luminous objects. As shown by \cite{patton08} and \cite{deravel08},
the merger rate increases significantly for minor mergers and less luminous galaxies. 
Therefore the larger value derived by \cite{wake08} is not necessarily in disagreement
with ours.

\cite{lin08} apply the same method as we do (dynamically close pairs) to galaxies in
the DEEP2 survey and derive an integrated merger rate of $24\%$ since $z < 1.2$ for
galaxies with $-21 < M_B < -19$. Assuming a flat evolution of the merger rate \citep{
lin08}, this is equivalent to $\sim 6\%$ per Gyr. Here, the different luminosity ranges
sampled, the bandpass difference, and the fact that dry mergers were selected by
morphology, rather than by colors and spectral features as we do, may play a role 
in explaining the difference.

\cite{bell06a} obtain an integrated dry merger rate of $\sim 80\%$ since $z=0.8$
for galaxies in the Combo-17 survey with $M_B < -20.5$, while applying the small
scale correlation function to the same data, \cite{bell06b} derive a merger rate 
of $4\%$ per Gyr at $0.4 < z < 0.8$. As noted by \cite{wake08}, the space density
of objects in these surveys is 20 times greater than ours, and therefore \cite{bell06a,
bell06b} are sampling considerably less luminous objects than we do. Given the
dependence of the merger rate on luminosity \citep{patton08,deravel08} this does not mean
that our results are in disagreement. In addition, \cite{khochfar08} show that
the above sample includes both wet and dry mergers, unlike ours where we confirm
by spectroscopy that the vast majority of the sample has LRG-type spectra \citep{
roseboom06}.

Finally, \cite{white07} measure an integrated dry merger rate of 30\% for LRGs in
the NDFWS survey between $0.5 < z < 0.9$. This is equivalent to a merger rate of
3.4\% per Gyr, but needs to be corrected for the higher mean redshift and different
space density than the 2SLAQ sample. \cite{wake08} estimate that this downward
revision is about a factor of 10, which places the result by \cite{white07} in
good agreement with ours.

In addition, there are several estimates of the merger rate for all galaxies using
asymmetries for the DEEP2 \citep{lotz08} and COSMOS surveys \citep{kampczyk07}, as
well as galaxies in the GOODS fields \citep{lopez09}, and pairs in the COSMOS survey
\citep{kartaltepe07}. The measured merger rate is approximately $2$--$4\%$ per Gyr
for the entire population of $L > L^*$ galaxies, of which dry mergers are only a
subset, which is in agreement with our measurement. 

Our result is also consistent with the more indirect merger fraction derived from the 
evolution of the red galaxy luminosity function. Note that, in this case, `growth' may 
take place via the transformation of blue galaxies into quiescent objects, moving on to 
the red sequence, but without (necessarily) any major merging. \cite{wake06} used the 2SLAQ 
data to show that the LRG luminosity function evolves passively at $z < 0.6$, a result 
confirmed by several other studies \citep{bundy06,caputi06,cimatti06,scarlata07}. For $\sim 
4 L^*$ LRGs \cite{brown07,brown08} and \cite{cool08} show that these galaxies evolve essentially 
passively at $z < 1$.

Some massive red mergers are observed in distant clusters and groups (e.g., \citealt{vandokkum99,
vandokkum01}). Using the pair fraction, \cite{vandokkum99} estimated that $\sim 50\%$ of massive 
red galaxies in the cluster MS1054-0321 at $z=0.82$ have undergone a merger to the present epoch. 
\cite{mcintosh07} calculate that, for groups and clusters in the SDSS at $z<0.12$, the merger rate 
for massive red galaxies is $2$--$9\%$ per Gyr. These are much higher merger rates than we measure, 
albeit in a much denser environment. Assuming that the merger rate scales with density, and since 
clusters are about 100 times denser than the field, the values derived by \cite{vandokkum99,
vandokkum01} and \cite{mcintosh07} are probably not fully inconsistent with our $\sim 0.3\%$ upper 
limit for field LRGs.

\subsection{Implications for Galaxy Formation}

At face value, our results are in considerable tension with models of galaxy formation in 
$\Lambda$CDM cosmologies where most massive red galaxies grow via dry mergers \citep{khochfar03,
delucia06,khochfar08}. \cite{khochfar08} use a semi-analytic model to model the quenching of
star formation and a standard merger tree, to predict the dry merger rate for galaxies above 
a characteristic mass (the mass for which star formation is shut off) at $z < 1$. For galaxies 
with $M > 6.3 \times 10^{10}$ M$_{\odot}$ \cite{khochfar08} predict a dry merger rate of $6 \times 
10^{-5}$ Mpc$^{-3}$ $h^{-3}$ Gyr$^{-1}$, and a typical merger fraction of about 10--20\% per Gyr,
almost independent of redshift. 

Our luminosity range of $-23 < M_r < -21.5$ translates into a mass range of 16 to 5 $\times
10^{10}$ M$_{\odot}$. The upper limit to the merger rate we derive ($<0.8 \times 10^{-5}$ 
Mpc$^{-3}$ $h^{-3}$ Gyr$^{-1}$, including the `worst case scenario' for systematic errors and 
assumuing all pairs merge on the shortest possible timescale) is at least a factor of 5 below
these predictions. It is therefore unlikely that dry mergers at $z < 0.7$ are important in the 
buildup of the red sequence \citep{bundy07,genel08b,bundy09}

On the other hand, the mass fraction in $\leq L^*$ galaxies on the red sequence appears to 
grow by about $50\%$ since $z=1$ \citep{bell04,borch06,faber07,cool08}. \cite{scarlata07} 
also note a deficit of fainter early-type galaxies in their COSMOS data at $z=0.7$. The
main mode of growth of the red sequence at $z < 1$ may be via cessation of star formation
and morphological evolution in lower mass galaxies \citep{bell04,faber07}.

We therefore favor a scenario where most massive galaxies are formed quasi-monolithically 
at high redshift and major mergers are relatively unimportant at $z < 1$ (e.g., \citealt{
bower06,naab07} -- cf., \citealt{conselice06} for an observational perspective). This
is consistent with recent CDM simulations, that downplay the role of major mergers
in galaxy formation over the past half of the Hubble time and favor minor mergers and
internal processes as drivers of galaxy evolution in the last $\sim 8$ Gyr (e.g., 
\citealt{cattaneo08,guo08,parry08}). However, even in these models the more massive LRGs 
should grow primarily by major dry mergers, a result that appears to be in some conflict 
with the slow growth observed for these galaxies \citep{brown07,brown08,cool08}.

It is important to distinguish between the merger rate of dark matter halos and
that of their galaxy tracers. Strictly speaking, theory can only predict the former,
while the latter depends, at least in part, on complex details of baryonic physics
and dynamical friction (cf. \citealt{berrier06}). Although our results appear to
favor a particular realization of $\Lambda$CDM models, interpretation must await
more detailed and realistic simulations of the stellar components of galaxies and
their fate during mergers and interactions.





\acknowledgments

We would like to thank Robert C. Nichol for some valuable comments on this paper.

We warmly thank all the present and former staff of the Anglo-Australian Observatory for their work 
in building and operating the 2dF facility. The 2SLAQ Survey is based on the observations made with 
the Anglo-Australian Telescope and for the SDSS. Funding for the SDSS and SDSS-II has been provided 
by the Alfred P. Sloan Foundation, the Participating Institutions, the National Science Foundation, 
the U.S. Department of Energy, the National Aeronautics and Space Administration, the Japanese 
Monbukagakusho, the Max Planck Society, and the Higher Education Funding Council for England. The 
SDSS Web Site is http://www.sdss.org/. 

The SDSS is managed by the Astrophysical Research Consortium for the Participating Institutions. 
The Participating Institutions are the American Museum of Natural History, Astrophysical Institute 
Potsdam, University of Basel, University of Cambridge, Case Western Reserve University, University 
of Chicago, Drexel University, Fermilab, the Institute for Advanced Study, the Japan Participation 
Group, Johns Hopkins University, the Joint Institute for Nuclear Astrophysics, the Kavli Institute 
for Particle Astrophysics and Cosmology, the Korean Scientist Group, the Chinese Academy of Sciences 
(LAMOST), Los Alamos National Laboratory, the Max-Planck-Institute for Astronomy (MPIA), the 
Max-Planck-Institute for Astrophysics (MPA), New Mexico State University, Ohio State University, 
University of Pittsburgh, University of Portsmouth, Princeton University, the United States Naval 
Observatory, and the University of Washington.



{\it Facilities:} \facility{AAT (2dF)}, \facility{SDSS}.

\clearpage

\end{document}